# Superconductivity above 200 K Discovered in Superhydrides of Calcium

Zhiwen Li [a,1,2], Xin He [a,1,2,3], Changling Zhang [a,1,2], Xiancheng Wang*[1,2], Sijia Zhang [1], Yating Jia [1], Shaomin Feng [1], Ke Lu [1,2], Jianfa Zhao [1,2], Jun Zhang [1,2], Baosen Min [1,2], Youwen Long [1,2,3], Richeng Yu [1,2], Luhong Wang [4], Meiyan Ye [5], Zhanshuo Zhang [5], Vitali Prakapenka [6], Stella Chariton [6], Paul A. Ginsberg [7], Jay Bass [7], Shuhua Yuan [8], Haozhe Liu [8], Changqing Jin*[1,2,3]

[1] Beijing National Laboratory for Condensed Matter Physics, Institute of Physics, Chinese Academy of Sciences, Beijing 100190, China. [2] School of Physical Sciences, University of Chinese Academy of Sciences, Beijing 100190, China. [3] Songshan Lake Materials Laboratory, Dongguan 523808, China. [4] Harbin Institute of Technology, Harbin 130001, China. [5] Jilin University, Changchun, China. [6]Center for Advanced Radiations Sources, University of Chicago, Chicago, Illinois 60637, USA. [7]Department of Geology, University of Illinois at Urbana-Champaign, Urbana, Illinois, 61801, USA. [8] Center for High Pressure Science & Technology Advanced Research, Beijing 100094, China.

## Abstract

**Searching for superconductivity with $T_c$ near room temperature is of great interest both for fundamental science & many potential applications. Here we report the experimental discovery of superconductivity with maximum critical temperature ($T_c$) above 210 K in calcium superhydrides, the new alkali earth hydrides experimentally showing superconductivity above 200 K in addition to sulfur hydride & rare-earth hydride system. The materials are synthesized at the synergetic conditions of 160~190 GPa and ~2000 K using diamond anvil cell combined with in-situ laser heating technique. The superconductivity was studied through in-situ high pressure electric conductance measurements in an applied magnetic field for the sample quenched from high temperature while maintained at high pressures. The upper critical field Hc(0) was estimated to be ~268 T while the GL coherent length is ~11 Å. The in-situ synchrotron X-ray diffraction measurements suggest that the synthesized calcium hydrides are primarily composed of $CaH_6$ while there may also exist other calcium hydrides with different hydrogen contents.**

# Introduction

Superhydrides have received growing attentions recently because of high temperature superconductivity (SC)[1]. It was suggested that superhydrides are likely to be a step toward metallic hydrogen since the "chemically precompressed" hydrogen in the superhydrides can decrease the critical pressure required for the metallization[1-3]. Because of high phonon frequencies and strong electron phonon coupling, high temperature SC was suggested in hydrogen rich materials. Sulfur hydride is one of example in which the high $T_c$ SC was reported both theoretically & experimentally at Megabar level pressures[4,5]. In addition, alkali-earth, rare-earth and other superhydrides are theoretically suggested to possibly possess high temperature SC at high pressures[6-11]. Those metal superhydrides favor to form hydrogen rich coordination such as clathrate-like cage structures at high pressure where the metal ions are located in the center of the H clathrates and act as carrier donor. The occupation of the unfilled anti-bonding σ* orbitals of the $H_2$ molecules by the electrons from the metal ions would enhance to dissociate the $H_2$ molecules hence form H clathrates. Since the discovery of SC in sulfur hydrides, other superhydrides have been experimentally reported to have SC, including $LaH_{10}$[11-14], $YH_9$[8,15], $CeH_9$[16], $ThH_{10}$[17], Zirconium polyhydrides[18], Tin hydrides[19], $YH_4$[20] and $LuH_3$[21] etc..

Although alkali-earth superhydrides were predicted to have high temperature SC, so far there are no experimental observations for SC above 200 K, in spite of SC with $T_c$ ~20 K was reported in barium hydrides[22]. Here, we report the experimental discovery of high $T_c$ SC in calcium superhydrides synthesized at high pressure and high temperature conditions. The SC with $T_c$ ~210 K at 160 ~190 GPa was observed. Hence in addition to sulfur hydrides and rare-earth hydrides, this is a new alkali-earth hydride superconductor with $T_c$ above 200 K.

## Results & Discussions

**High Temperature Superconductivity.** Fig. 1(a) displays the image of the sample assembly. The dotted red square is the sample shape and the four gray dotted regions are the inner electrodes composed of Pt. Fig. 1(b) shows the measurements of resistance as function of temperature at 160 GPa for Sample A. The resistance curves almost coincide for both the cooling and warming cycles. The resistance drops at around 210 K and approaches to zero at low temperature, suggesting that the superconducting like transitions take place in the calcium superhydrides. Fig. 2 shows the resistance measurements by application of magnetic fields for Sample A. It is seen that the whole transition curves shift to low temperature with the critical temperature gradually suppressed by the applied field, consistent with the nature of superconductivity. The inset of Fig. 1(b) displays an enlarged view of the derivative of resistance over temperature around the transition region, from which the $T_c^{onset}$ of 210 K is estimated referring to right side upturn point. The experimentally observed $T_c$ value of 210 K is comparable to the theoretical predictions for the calcium hydride $CaH_6$ that crystalizes in a cubic structure with the space group of *Im*-3*m*[6].

In addition to the first transition at ~210 K, there are also several drop kinks at ~180 K and ~160 K as shown in Fig. 1(b). These drop kinks imply multistep superconducting transitions that are supported by the broaden transition curves shown in Fig.2. In the context of theoretical calculation, the calcium hydrides such as $CaH_6$, $CaH_{12}$ *etc* are predicted to have high $T_c$ SC at high pressure[2, 6]. Since near the experimental pressure of 160 GPa, $CaH_4$, $CaH_6$ and $CaH_{12}$ are predicted to be stable, the stoichiometric ratio of the synthesized superconducting hydrides $CaH_x$ should be mainly determined by the hydrogen content released during the laser heating process. Hence it is likely that these multistep transitions should be attributed to calcium hydrides with different hydrogen amounts, which are probably due to inhomogeneous hydrogen distribution.

To increase homogeneity of the generated hydrogen, more homogenous scanning heating manner controlled by computer was adopted. We performed several sample synthesis to achieve a single superconducting phase transition by optimizing pressure and temperature conditions. Fig. 3(a) presents the resistance for Samples B and C as function of pressure measured at 185 GPa, where a sharp superconducting transition are achieved. The SC is supported by the measurements at applied magnetic field as shown in Fig. 3(b). For the Sample D, the superconducting transition is shown in Supplementary Fig. 1. Thus, our results show the experimental discovery of high temperature alkali earth superhydride with $T_c$ exceeding 200 K, joining the 200K above superhydrides of $SH_3$ and $LaH_{10}$ and $YH_9$. Since hydrogen is the lightest element, superhydrides are expected to have high Debye temperature and thus possible high $T_c$ SC according to the prediction by conventional BCS theory. In addition, for the reported binary superhydrides with high $T_c$ SC, it was found that most of elemental metals are on the boundary of $s$ and $d$ blocks in the elemental period table[2], such as La in $LaH_{10}$ and Y in $YH_9$. For these elemental metals, electrons can be transferred from filled $s$ orbital to $d$ orbital under high pressure due to the small energy difference between these orbitals, which usually increases the structural instability and enhances the electron phonon coupling strength and thus leads to possible SC with high $T_c$ as well [2, 23, 24]. The Ca lattice in $CaH_6$ predicted in Ref.6 is highly compressed under high pressure, which suggests the possible large electron phonon coupling strength caused by the electron $s$-$d$ orbital transfer. Therefore, the combination of the high Debye temperature and the large electron phonon coupling strength makes it possible to reach high $T_c$ SC in calcium superhydrides.

**Physics Parameters Measurements.** To estimate the upper critical field $H_{c2}(0)$, the superconductivity as a function of magnetic field was investigated based on Sample C. The $T_c$ values are determined with the criteria of the $T_c^{onset}$, the temperature at 90% and 50% of normal state resistance ($T_c^{90\%}$ and $T_c^{50\%}$), respectively. By linear fitting the $H_{c2}(T)$, the slope of $dH_{c2}/dT|_{Tc}$ was obtained to be -1.976 T/K for $T_c^{onset}$, -1.726 T/K for $T_c^{90\%}$ and -1.389 T/K for the $T_c^{50\%}$. Using the Werthamer-Helfand-Hohenberg (WHH) formula of $\mu_0 H_{c2}(T) = -0.69 \times dH_{c2}/dT|_{Tc} \times T_c$, the $H_{c2}(0)$ value was calculated to be 268 T, 226 T and 180 T for the criteria of $T_c^{onset}$, $T_c^{90\%}$ and $T_c^{50\%}$, respectively. The Ginzburg Landau (GL) formula of $\mu_0 H_{c2}(T) = \mu_0 H_{c2}(0)(1 - (T/T_c)^2)$ is also used to estimate the upper critical field at zero temperature. As shown in Fig. 4 the fitting of the $\mu_0 H_{c2}(T)$ by GL formula gives a value of $H_{c2}(0)$ ranging from 131 T to 196 T using criteria of $T_c^{onset}$, $T_c^{90\%}$ and $T_c^{50\%}$, respectively. These are comparable to values reported for LaH$_{10}$ [12]. Using the obtained value of $H_{c2}(0)$=131~268 T, one can roughly estimate the GL coherence length $\xi$ to be 11~16 Å via the equation of $\mu_0 H_{c2}(0) = \Phi_0/2\pi\xi^2$, where $\Phi_0 = 2.067 \times 10^{-15}$ Web is the magnetic flux quantum.

**Structure Characterizations.** The possible superconducting phases are investigated by high pressure X-ray diffraction experiments. Supplementary Fig. 2(a) shows the room temperature X-ray diffraction pattern for a synthesized sample, where the $Im\bar{3}m$ phase of CaH$_6$ was indexed. Supplementary Fig. 2(b) shows the X-ray diffraction patterns measured at various high pressures starting from the synthesized pressure up to 243 GPa where the anvils were broken. The pressure dependence of unit cell volume of the CaH$_6$ is presented in Fig. 5. By using the Birch–Murnaghan equation (1),

$$P(\text{GPa}) = \frac{3}{2} \times B_0 \left[ \left(\frac{V_0}{V}\right)^{\frac{7}{3}} - \left(\frac{V_0}{V}\right)^{\frac{5}{3}} \right] \times \left\{ 1 - \left(3 - \frac{3}{4} \times B_0'\right) \times \left[ \left(\frac{V_0}{V}\right)^{\frac{2}{3}} - 1 \right] \right\} \quad (1),$$

where $B_0$ is the bulk modulus and $B_0$' is the pressure derivative bulk modulus, the *P-V* data are fitted and $B_0$ is estimated to be about 221 GPa at $B_0$'= 3. From the fit of equation of state the lattice constant *a* at 150 GPa is estimated to be 3.308 Å, which is close to that predicted by theoretical calculation. This suggests that the superconducting sample contains $Im\bar{3}m$ phase of $CaH_6$. However, other phase(s) might be mixed since there are some unknown weak peaks in the diffraction patterns. We became aware during preparing the paper that an independent work by Ma *et al*. was carried with the similar results[25].

In summary high Tc calcium superhydrides superconductors are discovered with critical temperature Tc above 210 K at pressures ~ 160GPa. The upper critical field Hc(0) was estimated to be ~268 T that corresponds to ~11 Å coherent length in GL model for the calcium superhydride superconductor. The results show that calcium superhydrides are new type hydrogen rich superconducting compounds with transition temperature above 200 K in addition to sulfur hydride & rare-earth hydride system.

## Methods

**High Pressure Assembly.** Diamond anvil cell techniques are used for both synthesis & measurements of the calcium hydrides. The diamond anvils are of optical pure quality with culet diameter of 100 μm beveled to 300 μm for the experiments. The electrodes for resistance measurements are embedded in the sample chamber before high pressure & high temperature synthesis. The experiments are conducted in two steps: (I) synthesis of the calcium hydrides at high pressure & high temperature; (II) the follow up measurements of SC for the sample quenched from high temperature but remained at the same pressure. The T301 stainless steel served as gaskets that are prepressed from 250 μm to ~20 μm in thickness at about 20 GPa. A hole of 300 μm in diameter was drilled at the center of imprint, and was filled with fine insulating cubic boron nitride (cBN) powder mixed with epoxy that was pressed with anvils to form a solid layer of ~15 μm in thickness. A hole of 70 μm in diameter was drilled at the center of the solid cBN layer to serve as the sample chamber. The Pt foils with a thickness of 0.5 μm were deposited on the surface of the culet as the inner electrodes. Gold wire was attached to the Pt foil to serve as the outside electrodes. High purity calcium metal (99.9%) was adopted as precursor material: a 2 μm thick calcium specimen with dimensions 30 μm * 30 μm was stacked on the inner electrodes. A flake of ammonia borane was inserted into the sample chamber to act as a hydrogen source (hydrodizer) while also serving as the pressure transmitting medium[14]. The sample loading as well as electrodes deposition were conducted in a glove box filled by flowing Ar gas with 1 ppm less trace water or oxygen to avoid moisture or contamination. The diamond anvil cell was then clamped in the glove box before the measurements. The described setup followed the ATHENA procedure reported Ref.26.

**Sample Synthesis and Resistance Measurements.** A piston cylinder type anvil cell composed of BeCu was adopted in the experiments. The sample was first pressed to 160~190 GPa followed by heating at ~2000 K for ~6 min. The heating was performed with a pulse laser beam of 1064 nm wavelength. The laser had a focused spot size 5 μm in diameter with a power of 20 W. The heating temperature was estimated by fitting the black body irradiation spectra. A computer controlled scanning heating method was adopted to increase the homogeneity of the generated hydrogen. The heated sample was then quenched from high temperature while held at high pressure. The synthesized sample compressed within diamond anvil cell was put into a MagLab system to perform the electric conductivity measurements using a Van der Pauw method with 1 mA applied current. The MagLab system can provide synergetic extreme environments with temperatures from 300 K to 1.5 K and a magnetic field up to 9 T[27,28]. Pressure was calibrated by the shift of the first order Raman edge frequency from the cutlet of diamond[29].

**High Pressure x-ray Diffraction Measurements.** The synchrotron X-ray diffraction experiments in the diamond anvil cell are carried out at GSECARS of the Advanced Photon Source at the Argonne National Laboratory. X-ray with $\lambda$ = 0.3344 Å are focused down to ~3 um diameter spot on the sample. A symmetric diamond anvil cell was used to generate pressure with beveled diamond anvils (100/300um) and a rhenium gasket indented to 25 μm. Samples are loaded into the pressure chamber without cBN insulating layer. Calcium hydrides are synthesized in situ at 190 GPa with laser heating at 2000 K. The sample pressure in the synchrotron X-ray diffraction experiments was determined by both the shift of Raman edge frequency from the cutlet of diamond and the equation of state for rhenium. The X-ray diffraction images are converted to one dimensional diffraction data with Dioptas[30].

**Acknowledgments:**

The work is supported by NSF, MOST & CAS of China through research projects. Portions of this work were performed at GeoSoilEnviroCARS (The University of Chicago, Sector 13), Advanced Photon Source (APS), Argonne National Laboratory. GeoSoilEnviroCARS is supported by the National Science Foundation–Earth Sciences (EAR–1634415) and Department of Energy-GeoSciences (DE-FG02-94ER14466). This research used resources of the Advanced Photon Source, a U.S. Department of Energy (DOE) Office of Science User Facility operated for the DOE Office of Science by Argonne National Laboratory under Contract No. DE-AC02-06CH11357. We are grateful to Prof. J.G. Cheng, J.P. Hu & L. Yu for the discussions. We thank Prof. T. Xiang, B.G. Shen, & Z.X. Zhao for the consistent encouragements.

## Author Contributions

Research Design, Coordination & Supervision: C.Q.J.; high pressure synthesis and in situ resistance measurements: Z.W. L., X.H., C.L.Z., S.J.Z., Y.T.J., S.M.F., Y.W.L.,X.C.W., J.Z., B.S. M., R.C.Y, J.F.Z., C.Q.J; in situ synchrotron experiments: L. H. W., M.Y.Y., Z.S.Z., V.B.P., S.C., P. A. G., J. B., S. H. Y., H. Z. L., manuscript writing: Z.W.L., X.C.W. and C.Q.J. All authors contributed to the discussions.

## Competing interests:

The authors declare no competing interests.

**Figure Captions**:

Fig. 1: The superconductivity measurements. (a) The image of the specimen assembly. (b) The temperature dependence of resistance $R(T)$ measured at 160 GPa and zero magnetic field for Sample A. The multiple steps in the transition region imply there might be phases with different hydrogen amount in the synthesized calcium superhydrides. The inset is the enlarged view of the derivative of resistance over temperature($dR/dT$) where the ~ 210 K $T_c^{onset}$ is determined referring to the upturn point at the right side.

Fig. 2: The superconducting transitions with the applied magnetic fields. The resistance as a function of temperature measured at different magnetic fields for Sample A. The arrows are attributed to the phases with different hydrogen amount in the calcium superhydrides. The inset is the enlarged view of the resistance curves around the transition region of the 210 K phase.

Fig. 3: Evolution of superconducting transitions as functions of magnetic fields. (a) The temperature dependence of resistance measured for Samples B and C at 185 GPa showing one step transition; (b) Resistance curves measured under different magnetic field for Sample C.

Fig. 4: The superconducting parameters. The Ginzburg Landau fitting for the $H_{c2}(T)$ shown with the solid lines. The stars represent the $H_{c2}(0)$ values calculated via WHH model. The inset of Fig. 4 is the critical field Hc2(T) as a function of $T_c$ with the $T_c$ values determined by the criteria of the $T_c^{onset}$, 90% and 50% of normal state resistance, respectively.

Fig. 5: The structure characterizations. The pressure dependence of unit cell volume of the $CaH_6$. The red line is the fit to equation of stat

**Fig. 1**

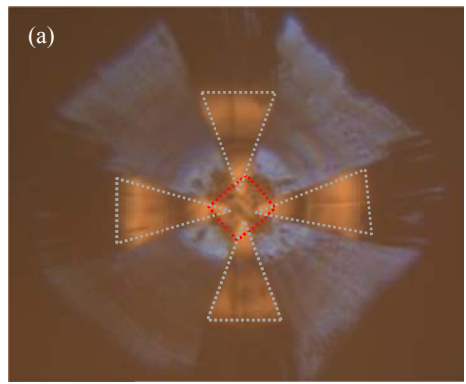

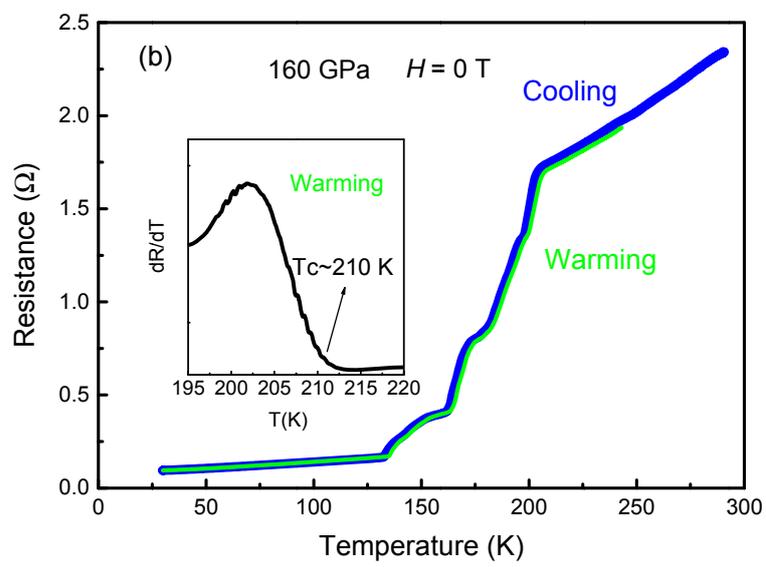

**Fig. 2**

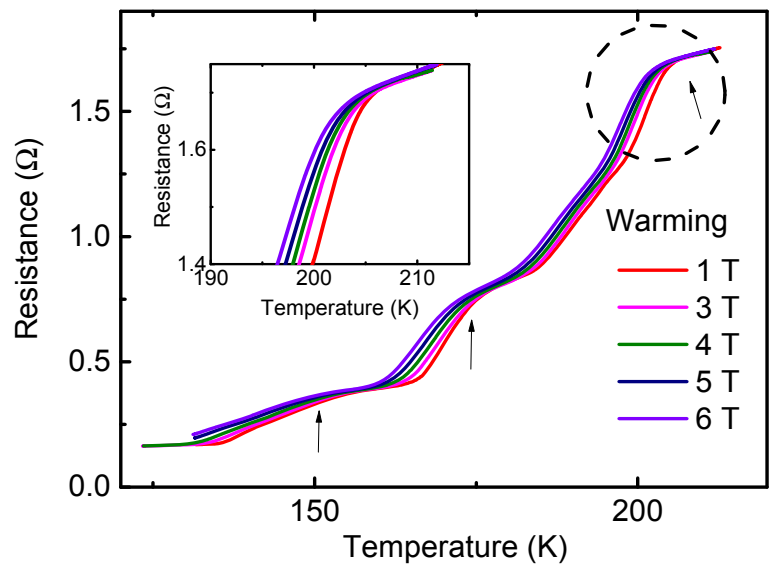

**Fig. 3**

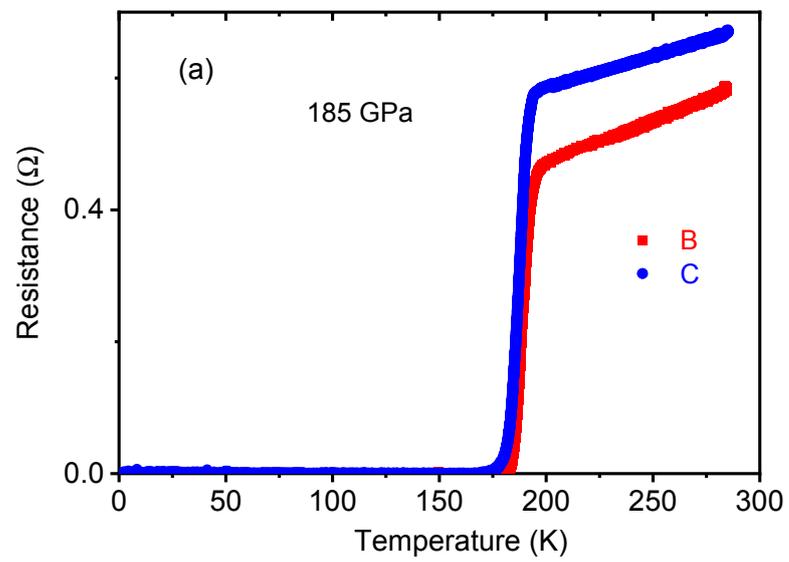

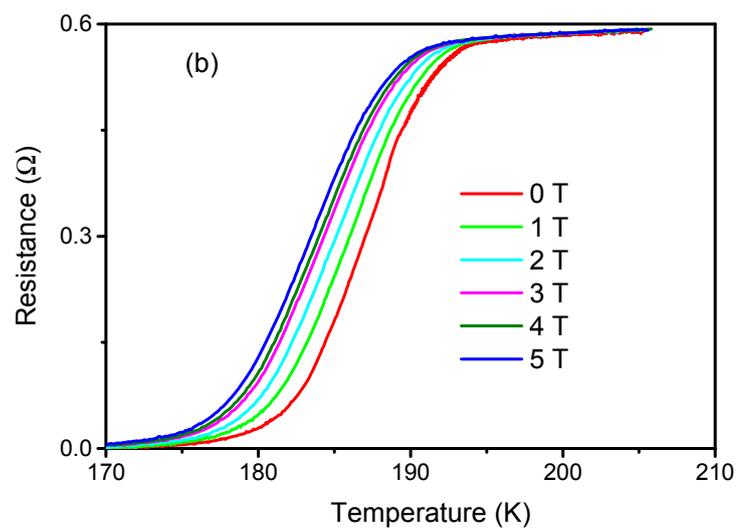

**Fig. 4**

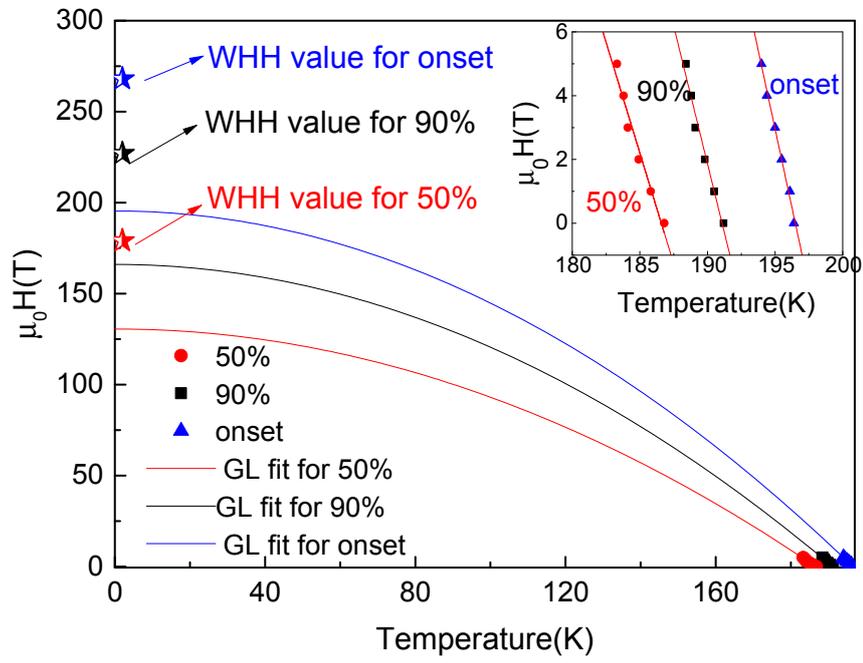

**Fig. 5**

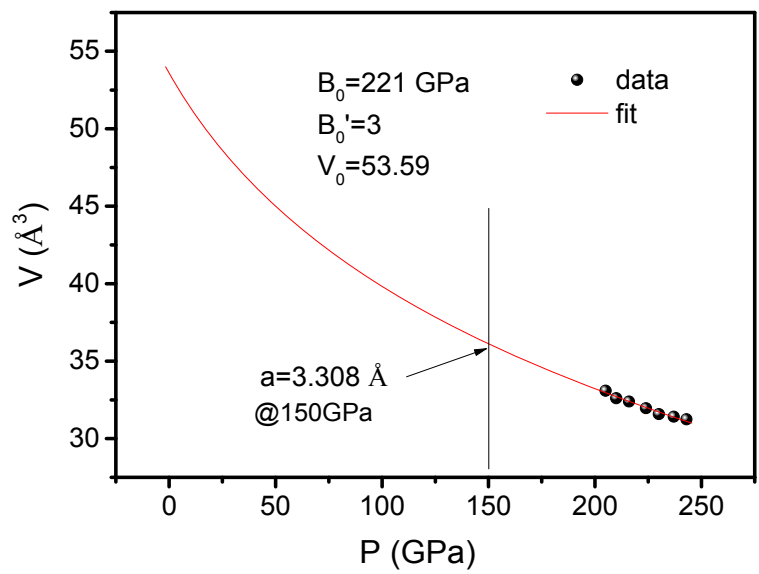